%% file: main.tex
\documentclass[sigconf]{acmart}
\pdfoutput=1 
\AtBeginDocument{%
  \providecommand\BibTeX{{%
    \normalfont B\kern-0.5em{\scshape i\kern-0.25em b}\kern-0.8em\TeX}}}

\usepackage{soul}
\usepackage{graphicx}
\usepackage{booktabs}
\usepackage{float}
\usepackage{overpic}
\usepackage{float}  
\usepackage{subfloat}
\usepackage{stfloats} 
\usepackage[skip=0.5\baselineskip]{caption}
\usepackage{bm}
\usepackage{amsmath}
\usepackage{booktabs}
\usepackage{multirow}
\usepackage{multicol}
\usepackage{enumitem}
\usepackage{subcaption}
\usepackage{threeparttable}
\usepackage{xspace}
\usepackage[normalem]{ulem}
\usepackage{algorithmicx,algorithm}
\usepackage{algpseudocode}
\usepackage{algorithm,algpseudocode}
\usepackage{marvosym}   
\usepackage{colortbl}   
\usepackage{tcolorbox}  
\tcbuselibrary{breakable}   
\usepackage{wasysym}    
\usepackage[figuresright]{rotating} 
\usepackage{pifont} 

\makeatletter
\def\@ACM@checkaffil{
    \if@ACM@instpresent\else
    \ClassWarningNoLine{\@classname}{No institution present for an affiliation}%
    \fi
    \if@ACM@citypresent\else
    \ClassWarningNoLine{\@classname}{No city present for an affiliation}%
    \fi
    \if@ACM@countrypresent\else
        \ClassWarningNoLine{\@classname}{No country present for an affiliation}%
    \fi
}
\makeatother

\usepackage{xspace}
\newcommand{\etal}{\emph{et al.}\xspace}
\newcommand{\eg}{\emph{e.g.,}\xspace}
\newcommand{\ie}{\emph{i.e.,}\xspace}

\copyrightyear{2025}
\acmYear{2025}
\setcopyright{acmlicensed}\acmConference[SIGIR '25]{Proceedings of the 48th
International ACM SIGIR Conference on Research and Development in
Information Retrieval}{July 13--18, 2025}{Padua, Italy}
\acmBooktitle{Proceedings of the 48th International ACM SIGIR Conference on
Research and Development in Information Retrieval (SIGIR '25), July 13--18,
2025, Padua, Italy}
\acmDOI{10.1145/3726302.3729911}
\acmISBN{979-8-4007-1592-1/2025/07}

\begin{document}
\begin{sloppypar}   
\title{Bridge the Domains: Large Language Models Enhanced Cross-domain Sequential Recommendation}

\author{Qidong Liu}
\affiliation{%
  \institution{Xi'an Jiaotong University \& \\ City University of Hong Kong}
  \city{Xi'an}
  \country{China}
}
\email{liuqidong@stu.xjtu.edu.cn}

\author{Xiangyu Zhao \Letter}
\thanks{\Letter \ \text{Corresponding authors}}
\affiliation{%
  \institution{City University of Hong Kong}
  \city{Hong Kong}
  \country{China}
}
\email{xianzhao@cityu.edu.hk}

\author{Yejing Wang}
\affiliation{%
  \institution{City University of Hong Kong}
  \city{Hong Kong}
  \country{China}
}
\email{yejing.wang@my.cityu.edu.hk}

\author{Zijian Zhang}
\affiliation{%
  \institution{Jilin University}
  \city{Changchun}
  \country{China}
}
\email{zhangzijian@jlu.edu.cn}

\author{Howard Zhong}
\affiliation{%
  \institution{City University of Hong Kong}
  \city{Hong Kong}
  \country{China}
}
\email{howzhong@cityu.edu.hk}

\author{Chong Chen}
\affiliation{%
  \institution{Tsinghua University}
  \city{Beijing}
  \country{China}
}
\email{cstchenc@163.com}

\author{Xiang Li}
\affiliation{%
  \institution{Nanyang Technological University}
  \city{Singapore}
  \country{Singapore}
}
\email{xiang002@e.ntu.edu.sg}

\author{Wei Huang}
\affiliation{%
  \institution{Independent Researcher}
  \city{Beijing}
  \country{China}
}
\email{hwdzyx@gmail.com}

\author{Feng Tian \Letter}
\affiliation{%
  \institution{Xi'an Jiaotong University}
  \city{Xi'an}
  \country{China}
}
\email{fengtian@mail.xjtu.edu.cn}

\renewcommand{\shortauthors}{Qidong Liu, \etal}

\begin{abstract}
  Cross-domain Sequential Recommendation (CDSR) aims to extract the preference from the user's historical interactions across various domains.
  Despite some progress in CDSR, two problems set the barrier for further advancements, \ie overlap dilemma and transition complexity. The former means existing CDSR methods severely rely on users who own interactions on all domains to learn cross-domain item relationships, compromising the practicability.
  The latter refers to the difficulties in learning the complex transition patterns from the mixed behavior sequences. 
  With powerful representation and reasoning abilities, Large Language Models (LLMs) are promising to address these two problems by bridging the items and capturing the user's preferences from a semantic view.
  Therefore, we propose an LLMs Enhanced Cross-domain Sequential Recommendation model (\textbf{LLM4CDSR}).
  To obtain the semantic item relationships, we first propose an LLM-based unified representation module to represent items. Then, a trainable adapter with contrastive regularization is designed to adapt the CDSR task. 
  Besides, a hierarchical LLMs profiling module is designed to summarize user cross-domain preferences. 
  Finally, these two modules are integrated into the proposed tri-thread framework to derive recommendations.
  We have conducted extensive experiments on three public cross-domain datasets, validating the effectiveness of LLM4CDSR. 
  We have released the code online\footnote{https://github.com/Applied-Machine-Learning-Lab/LLM4CDSR-pytorch}.
\end{abstract}



\maketitle

\input{1Introduction}

\input{2Preliminary}

\input{3Method}

\input{4Experiment}

\input{5RelatedWork}

\input{6Conclusion}

\input{8Acknowledgement}

\bibliographystyle{ACM-Reference-Format}
\bibliography{main}


\end{sloppypar}
\end{document}

%% file: 1Introduction.tex
\section{Introduction}  \label{sec:intro}

Sequential recommender systems (SRS), predicting the next item according to the user's historical behaviors, have been widely applied to various fields.
Many efforts have been devoted to fabricating neural architectures to capture users' dynamic preferences from their interaction records, such as recurrent neural networks in GRU4Rec~\cite{hidasi2015session} and self-attention mechanisms~\cite{vaswani2017attention} in SASRec~\cite{kang2018self}. 
However, the data sparsity problem in recommender systems~\cite{qiu2022contrastive} remains a major challenge to the construction of effective SRS.

\begin{figure}[!t]
\centering
\includegraphics[width=0.9\linewidth]{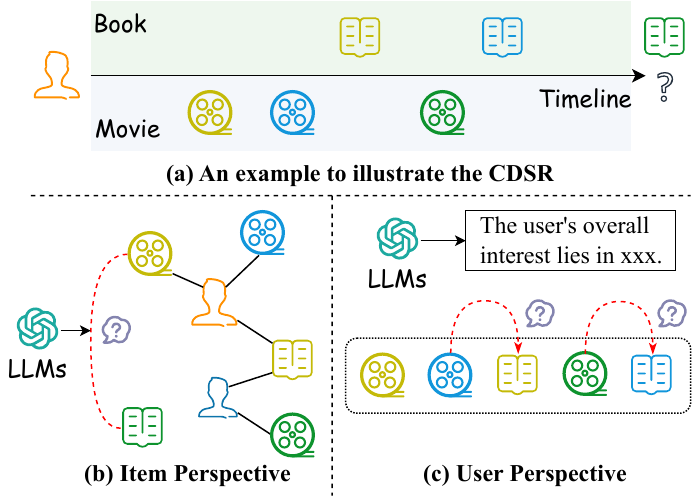}
\caption{The illustration of CDSR task and existing methods.}
\label{fig:preliminary}
\vspace{-5mm}
\end{figure}

Recently, the emergence of cross-domain sequential recommendation (CDSR) has helped alleviate the data sparsity problem by absorbing interactions from various domains~\cite{chen2024survey,cao2022contrastive,xu2024rethinking}. Specifically, CDSR organizes items from all domains and 
extracts transferable preferences across domains.
As the example in Figure~\ref{fig:preliminary}(a), interactions with period dramas (in green) can suggest similar preferences for historical books (in green), where we present this transferable pattern with the same color for cross-domain items.
Generally, how to capture such preferences from domains with distinct distributions is the key to CDSR~\cite{zang2022survey}, \ie bridging the domains. 
Current CDSR studies have explored this problem from two perspectives. 
\begin{itemize}[leftmargin=*]

    \item \textbf{i) Item Perspective}: This line of works~\cite{cao2022contrastive,wang2023unbiased} adopts graph neural networks (GNN)~\cite{wu2022graph} to establish the bridges of items from different domains. Taking the example in Figure~\ref{fig:preliminary}(b) for illustration, they build an interaction graph based on the records from all domains.
    Through a GNN, the relationship between the movie in blue and the book in yellow is built by the users who have interacted with both domains, called overlapping users.
    However, they often rely on these overlapping users to well-train the item embedding, denoted as \ding{182} \textbf{Overlap Dilemma}. Some emerging works, \eg AMID~\cite{xu2024rethinking}, have addressed this issue by grouping similar users to expand the overlap. 
    Nevertheless, they are still trapped in co-occurrence, which is severely affected by overlapping users. As the example in Figure~\ref{fig:preliminary}(b), the relationship between the movie in yellow and the book in green is difficult to gain due to no connection between them in the graph. 

    \item \textbf{ii) User Perspective}: Other CDSR studies~\cite{ma2024triple,park2024pacer} target capturing fine-grained user preference from mixed interaction sequences (including interactions from all domains). For instance, TriCDR~\cite{ma2024triple} proposes multiple contrastive objectives for nuanced global user representations.
    Nevertheless, the nature of complex cross-domain sequences makes it difficult to understand the transition pattern only from a collaborative view, \ie \ding{183} \textbf{Transition Complexity}.
    Taking the example in Figure~\ref{fig:preliminary}(c) for demonstration, it poses a difficulty in extracting the interest shifts between the alternate book and movie interactions.
    
\end{itemize}

Large Language Models (LLMs) have shown powerful abilities in representation and reasoning~\cite{zhao2023survey}. Thus, LLMs are promising to \textit{bridge} the semantic relationship between items~\cite{liu2024llm,hu2024enhancing}, \eg the connection between the movie in yellow and the book in green regardless of co-presence (Figure~\ref{fig:preliminary}(b)).
Since such connections are not affected by co-occurrence, \ie no need for overlapping users, LLMs can help alleviate the \ding{182} Overlap Dilemma. 
Besides, powerful LLMs can derive the overall preference of users by \textit{bridging} mixed interactions directly, as shown in Figure~\ref{fig:preliminary}(c). Such overall preferences can ease analyzing the fine-grained interest shift between interactions from different domains, thus addressing the \ding{183} Transition Complexity issue. 
In a word, LLMs are promising to establish better \textit{bridges} for CDSR from both item and user perspectives.
However, there are still two challenges that hinder the direct adaptation of LLMs to CDSR. 
\textbf{i) Inadaptability}. LLMs are trained for general purposes~\cite{touvron2023llama}, but not specialized for recommendation. Thus, the results generated from LLMs may not be suitable for the CDSR task directly.     
\textbf{ii) Lengthy Prompt}. Previous studies~\cite{geng2024breaking} have proven LLMs are not skilled in understanding prompts with too many words. Cross-domain sequences are often longer due to more domains and thus lead to lengthy prompts, which may cause the sub-optimal performance of LLMs for user profiling.    

Inspired by the \textit{bridge} idea discussed above, we propose a Large Language Models Enhanced Cross-domain Sequential Recommendation method (\textbf{LLM4CDSR}). 
From the item perspective, we devise a unified LLMs representation module to derive a set of item embeddings for various domains.
To conquer the inadaptability challenge, we design a trainable adapter and a contrastive regularization.
In terms of the user perspective, a hierarchical LLMs profiling module is designed, which can derive users' overall preferences. 
Targeting the lengthy prompt problem, we partition the mixed sequence into several parts and then apply an LLM-based summarizer to obtain the comprehensive user profile. 
Notably, since the LLMs embeddings and user profiles can be cached, LLM4CDSR eliminates LLMs inference while serving, guaranteeing efficiency.
The contributions of this paper are as follows: 
\begin{itemize}[leftmargin=*]
    \item In this paper, we design the LLM4CDSR method to address the overlap dilemma and transition complexity issues in CDSR.
    \item To fill the inadaptability gap between LLMs and the CDSR task, we design a unified LLMs representation module. To avoid lengthy prompts, we propose a hierarchical LLMs profile module.
    \item Experiments on three datasets validate that LLM4CDSR outperforms existing CDSR and LLM-based baselines consistently.
\end{itemize}


%% file: 2Preliminary.tex
\section{Problem Definition}

In this section, the problem definition of the CDSR task is given. Specifically, we focus on the two-domain situation. Let $\mathcal{U}=\{u_1,u_2,\dots,u_{\mid \mathcal{U}\mid}\}$ denote the user set, while $\mathcal{A}=\{a_1,a_2,\dots,a_{\mid \mathcal{A}\mid}\}$ and $\mathcal{B}=\{b_1,b_2,\dots,b_{\mid \mathcal{B}\mid}\}$ are item sets of domain $A$ and $B$, respectively. $|\mathcal{U}|$, $|\mathcal{A}|$ and $|\mathcal{B}|$ represent the number of users, items of domain $A$ and items of domain $B$.
For the general sequential recommendation, user interactions are ordered by the timeline within each domain, \ie $\mathcal{S}^A_u = \{a^{(u)}_1,a^{(u)}_2,\dots,a^{(u)}_{n_A}\}$ and $\mathcal{S}^B_u = \{b^{(u)}_1,b^{(u)}_2,\dots,b^{(u)}_{n_B}\}$, where $n_A$ and $n_B$ are the lengths of interaction sequences in domain $A$ and $B$ of user $u$. For simplicity, we eliminate the superscript and subscript for the user in the following notations. 
To utilize the data from both domains, CDSR organizes all interactions into one sequence, \ie $\tilde{\mathcal{S}} = \{v_1,v_2,\dots,v_{n_u}\}$, where $v_i \in \mathcal{A} \cup \mathcal{B}$ can be any item of domain $A$ or $B$. $n_u=n_A+n_B$ is the total number of the user's historical records. Then, given the interaction sequences in each domain, CDSR transforms the general SRS task across two domains into the following formulations:
\begin{equation}
\begin{aligned}
    \arg \max_{v_i \in \mathcal{A}} P(v_{n_u+1} & =v_i \mid \tilde{\mathcal{S}}, \mathcal{S}^A, \mathcal{S}^B) \\ 
    \arg \max_{v_i \in \mathcal{B}} P(v_{n_u+1} & =v_i \mid \tilde{\mathcal{S}}, \mathcal{S}^A, \mathcal{S}^B)
\end{aligned}
\end{equation}

%% file: 3Method.tex
\section{Method}


\begin{figure*}[t]
\centering
\includegraphics[width=0.8\linewidth]{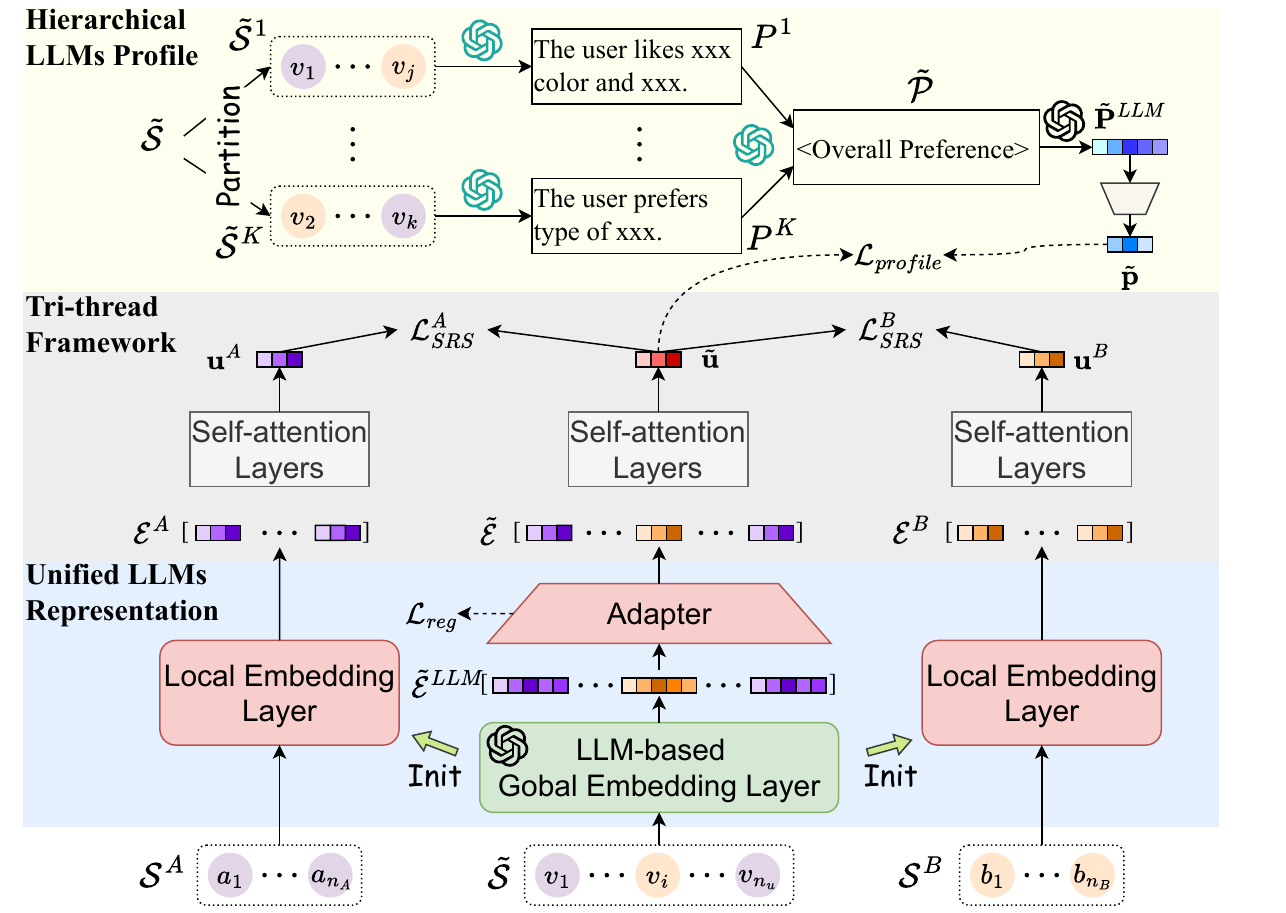}
\caption{The overview of the proposed LLM4CDSR.}
\label{fig:framework}
\vspace{-2mm}
\end{figure*}

\subsection{Overview} \label{sec:method_overview}
In Figure~\ref{fig:framework}, we show the overview of the proposed LLM4CDSR. It absorbs the historical behaviors in each domain, \ie $\mathcal{S}^A$ and $\mathcal{S}^B$, and the mixed interaction sequence of both domains, \ie $\tilde{\mathcal{S}}$. 
For capturing the local preference specified for each domain and the global preference shared by both domains, we propose a \textbf{Tri-thread Framework}, introduced in Section~\ref{sec:method_framework}. Among the framework, two threads encode the embedding sequences of each domain ($\mathcal{E}^A$ and $\mathcal{E}^B$), while the other is for the mixed interaction embedding sequence, \ie $\tilde{\mathcal{E}}$. The derived local user representations ($\mathbf{u}^A$, $\mathbf{u}^B$) along with the global user representation $\tilde{\mathbf{u}}$ will be utilized to give out the final recommendations for both domains.

From the item perspective, we elaborate a \textbf{Unified LLMs Representation} module to get the item embeddings, detailed in Section~\ref{sec:method_item}. To bridge the various distributions of domains, we propose adopting LLMs to generate semantic embeddings for all items. Then, a trainable adapter is designed to transform them into the final item embeddings. Besides, we initialize the local embedding layers by LLMs embeddings to alleviate difficulties in training.
From the user perspective, in Section~\ref{sec:method_user}, we devise a \textbf{Hierarchical LLMs Profile} module to understand users' cross-domain interests from an overall view. In this module, we first design a partition mechanism to break the mixed sequence into several parts, alleviating the lengthy prompts. Then, the preference in each part will be summarized and gathered by LLMs. The generated textual preference will be encoded by LLMs and the representation $\tilde{\mathbf{P}}^{LLM}$ is injected into the LLM4CDSR by alignment loss $\mathcal{L}_{profile}$ during the training process.

\subsection{Tri-thread Framework} \label{sec:method_framework}

For the cross-domain sequential recommendation task, the interests in each specified domain (\ie Local Preference) and the ones in both domains (\ie Global Preference) are equally vital~\cite{chen2024survey}. 
Generally, how to extract these two types of preferences and combine them effectively is one core challenge for CDSR.
To face such a challenge, we propose a tri-thread framework, consisting of three separate preference extraction lines. 
For each domain, the subsequent training and inference are based on domain-specific local threads and a global thread shared by both domains.
We take the CDSR for domain $A$ as an illustration, while a similar process is for domain $B$.

To capture both local and global preferences, the interaction sequence $\mathcal{S}^A$ and mixed sequence $\tilde{\mathcal{S}}$ are input to our model. 
In general, the identity sequences are transformed into embedding sequences as the first step, learning connections between items~\cite{zhao2023embedding}. We follow such a pattern, but propose a novel unified LLM representation module, which will be introduced in Section~\ref{sec:method_item}. 
The derived embedding sequences are denoted as $\mathcal{E}^A=\{\mathbf{e}^A_1, \mathbf{e}^A_2, \dots \mathbf{e}^A_{n_A}\}$ and $\tilde{\mathcal{E}}=\{\tilde{\mathbf{e}}_1, \tilde{\mathbf{e}}_2, \dots \tilde{\mathbf{e}}_{n_u}\}$, where $\mathbf{e}^A_i, \tilde{\mathbf{e}}_i \in \mathbb{R}^{d}$ are local and global item embeddings, respectively. $d$ is the size of embeddings. 
Then, the self-attention layers~\cite{vaswani2017attention} are imposed to $\tilde{\mathcal{E}}$ and $\mathcal{E}^A$ to capture the global and local preferences, encoding the user representations $\tilde{\mathbf{u}}\in \mathbb{R}^{d}$ and $\mathbf{u}^A \in \mathbb{R}^{d}$. The process can be formulated as follows:
\begin{equation} \label{eq:user}
    \tilde{\mathbf{u}} = f_{\tilde{\theta}}(\tilde{\mathcal{E}}), \ \mathbf{u}^A = f_{\theta_A}(\mathcal{E}_A)
\end{equation}
\noindent where $f_{\tilde{\theta}}(\cdot), f_{\theta_A}(\cdot)$ represent the self-attention layers.

For the final predictions, we devise a logit fusion to utilize both local and global user preferences. Specifically, we calculate the probability that the item $i$ will be preferred by the user $u$ as:
\begin{equation}    \label{eq:predict}
    P(v_{n_u+1}=v_i | \tilde{\mathcal{S}}, \mathcal{S}^A, v_i\in \mathcal{A}) = [\tilde{\mathbf{e}}_i:\mathbf{e}_i^A]^T [\tilde{\mathbf{u}}:\mathbf{u}^A]
\end{equation}
\noindent where $[:]$ means the vector concatenation operation. According to the prediction probability, we formulate the loss function to optimize the framework as follows:
\begin{equation}    \label{eq:loss_srs}
    \mathcal{L}_{SRS}^A = -\sum_{u\in\mathcal{U}}\sum_{j=1}^{n_A} \log \sigma(P(v_{j+1}=v^+)-P(v_{j+1}=v^-))
\end{equation}
\noindent Here, $v^+$ and $v^-$ represent the ground truth and randomly sampled negative items, respectively. 
For the other domain, $\mathcal{L}_{SRS}^B$ can be obtained by replacing `A' with `B' in Equation~\eqref{eq:user}-\eqref{eq:loss_srs}.

\subsection{Unified LLMs Representation} \label{sec:method_item}

The previous CDSR works often adopt Graph Neural Networks~\cite{cao2022contrastive} or normal embedding layers~\cite{ma2024triple,xu2024rethinking} to learn collaborative connections. However, they often face the overlap dilemma, as we detailed in Section~\ref{sec:intro}. To overcome this, we propose a unified LLMs representation module to learn semantic relationships between items, regardless of the overlapping interactions. 
Such relationships are reflected by the item embeddings, so the proposed module mainly transforms the item identity $v_i$ into the corresponding embedding $\mathbf{e}_i$. 
As visualized in Figure~\ref{fig:framework}, among three threads of LLM4CDSR, two of them are \textbf{Local Embedding} layers specified for the particular domain, and the other one is the \textbf{Global Embedding} layer shared by both domains.
Besides, we propose a \textbf{Contrastive Regularization} to facilitate the training of global item embeddings.

\subsubsection{\textbf{Global Embedding}} \label{sec:method_global}
This domain-shared thread aims to extract cross-domain item connections from the mixed interaction sequence. Previous works often fall short of this due to the overlap dilemma. Inspired by the powerful semantic understanding abilities of LLMs~\cite{zhao2023survey}, we propose utilizing LLMs to derive semantic item embeddings from the abundant textual contents with item attributes, such as ``Title''. These semantic embeddings are naturally unified and do not rely on overlapping users. Specifically, we construct the following item prompts:


\begin{tcolorbox}[
        colframe=brown,
        width=1\linewidth,
        arc=1mm, 
        auto outer arc,
        title={Item Prompt Template (Cloth)},
        breakable,]
        
        The cloth item has the following attributes: \\
        name is \ul{<TITLE>}; brand is \ul{<BRAND>}; rating is \ul{<DATE>}; price is \ul{<PRICE>}. \\
        The item has the following features: \ul{<FEATURE>}. \\
        The item has the following descriptions: \ul{<DESCRIPTION>}. 
        
\end{tcolorbox}

\noindent By filling the underlined places with corresponding attribute values, item prompts are then fed to LLMs. We subsequently take the hidden states from 
LLM's last transformer layer or adopt LLMs text embedding model\footnote{https://platform.openai.com/docs/guides/embeddings} to get item embeddings.
The LLMs embeddings of all items across two domains constitute the LLM-based global embedding layer, \ie $\mathbf{E}^{LLM} \in \mathbb{R}^{(\mid \mathcal{A} \mid + \mid \mathcal{B} \mid) \times d_{LLM}}$, containing the semantic relationships between items. $d_{LLM}$ is the dimension of LLMs embeddings. By taking the corresponding rows of embeddings from $\mathbf{E}^{LLM}$, we can get the LLMs embedding sequence $\tilde{\mathcal{E}}^{LLM}=\{\mathbf{e}^{LLM}_1,\mathbf{e}^{LLM}_2, \dots \mathbf{e}^{LLM}_{n_u}\}$ from input identity sequence $\tilde{\mathcal{S}}$.

However, the original LLMs embeddings are not suitable for recommendation tasks, because LLMs are often trained for general purposes~\cite{touvron2023llama}. Besides, the dimension of LLMs embeddings is often excessively large for SRS tasks. Thus, we propose a trainable adapter to transform the LLMs embeddings $\mathbf{e}^{LLM}_i$ into final item embeddings $\tilde{\mathbf{e}}_i$. The process is formulated as follows:
\begin{equation}
    \tilde{\mathbf{e}}_i=\mathbf{W}_1(\mathbf{W_2}\mathbf{e}^{LLM}_i + \mathbf{b}_2)+\mathbf{b}_1
\end{equation}
\noindent where $\mathbf{W}_1\in \mathbb{R}^{d\times \frac{d_{LLM}}{2}}$, $\mathbf{W}_2\in \mathbb{R}^{\frac{d_{LLM}}{2}\times d_{LLM}}$, $\mathbf{b}_1\in \mathbb{R}^{d}$, $\mathbf{b}_2\in \mathbb{R}^{\frac{d_{LLM}}{2}}$ are the trainable parameters of the adapter. By this transformation, we can get the item embedding sequence $\tilde{\mathcal{E}}$ for the shared thread. It is worth noting that the LLM-based global embedding layer $\mathbf{E}^{LLM}$ is frozen while training to maintain the semantics contained in original LLMs embeddings.

\subsubsection{\textbf{Local Embedding}}

The two local threads only capture users' preferences within the particular domain, so the local embeddings only construct intra-domain relationships between items. 
However, training local embeddings from scratch may suffer from the asynchronous and unbalanced training~\cite{amos2023never} under the tri-thread framework, where the global embeddings are from well-trained LLMs.
Thus, we suggest initializing the local embedding layer with the same LLMs embeddings. First, the embeddings of items in domain $A$ and $B$ are taken from $\mathbf{E}^{LLM}$, denoted as $\mathbf{E}^{LLM-A} \in \mathbb{R}^{(\mid \mathcal{A} \mid) \times d_{LLM}}$ and $\mathbf{E}^{LLM-B} \in \mathbb{R}^{(\mid \mathcal{B} \mid) \times d_{LLM}}$. Then, we conduct the Principal Component Analysis (PCA) algorithm~\cite{pearson1901liii} to get the local embeddings, \eg $\mathbf{E}^{LLM-A}\stackrel{PCA}{\longrightarrow}\mathbf{E}^A \in \mathbb{R}^{(\mid \mathcal{A} \mid) \times d}$. By taking the corresponding row from $\mathbf{E}^A$, we can transform $\mathcal{S}^A$ into the local item embedding sequence $\mathcal{E}^A$. The sequence $\mathcal{E}^B$ can be obtained via a similar operation to $\mathbf{E}^B$.

\subsubsection{\textbf{Contrastive Regularization}}
Previous steps have emphasized constructing unified semantic representations for global and local embeddings, but the domain-specific information is ignored. Thus, to better distinguish the items of various domains at the embedding level, we propose a contrastive regularization. 
Specifically, the designed regularization aims to pull the distance between co-occurred items in different domains and push the distance between irrelevant items.
In detail, we randomly take out two items of domain $A$ and $B$ from mixed sequence $\mathcal{S}$ to form a positive pair. After global embedding encoding, we can get the pair of embeddings $(\tilde{\mathbf{e}}_i^A, \tilde{\mathbf{e}}_i^B)$. Based on a batch of pairs, the in-batch contrastive loss function is formulated as follows:
\begin{equation} \label{eq:reg}
    \mathcal{L}^A_{reg}=-\frac{1}{Z} \sum_{i=1}^{Z} \log \frac{\exp({\rm cos}(\tilde{\mathbf{e}}_i^A, \tilde{\mathbf{e}}_i^B) / \gamma)}{\sum_{j=1}^Z \mathbb{I}_{[i \neq j]} \exp({\rm cos}(\tilde{\mathbf{e}}_i^A, \tilde{\mathbf{e}}_j^B) / \gamma)}
\end{equation}
\noindent where $Z$ is batch size, $\exp(\cdot)$ and ${\rm cos}(\cdot)$ represent the exponential and cosine similarity function, $\mathbb{I}_{[i \neq j]} \in \{0,1\}$ denotes the indicator function, $\gamma$ represents the temperature coefficient. 
In this formula, the item from domain $A$ is the anchor item. Symmetrically, exchanging the positions of $\tilde{\mathbf{e}}_i^A$ and $\tilde{\mathbf{e}}_i^B$ in Equation~\eqref{eq:reg} leads to the regularization loss with items from domain $B$ as the anchor, denoted as $\mathcal{L}_{reg}^B$. Thus, the ultimate contrastive regularization is the sum of two sides, \ie $\mathcal{L}_{reg} = \mathcal{L}_{reg}^A+\mathcal{L}_{reg}^B$.

\subsection{Hierarchical LLMs Profile} \label{sec:method_user}

LLMs have been proven able to understand and summarize users' preferences by textual historical interactions~\cite{bao2023tallrec,ren2024representation}. Motivated by this, we propose to utilize LLMs to address the user perspective transition complexity issue. However, the mixed interaction sequences are relatively long, leading to lengthy prompts. To fully inspire the potential of LLMs, we propose a hierarchical LLMs profiling method, which consists of two necessary steps, \ie \textbf{Partition} and \textbf{Profiling}.
Besides, we introduce \textbf{Alignment} to integrate the derived profile into our LLM4CDSR.

\subsubsection{\textbf{Partition}}
Since different sub-sequences of similar items may indicate diverse user interests~\cite{cen2020controllable},
similar items should be deposited into one sub-sequence.
To facilitate the following profiling step, where LLMs are used to hierarchically summarize the partitioned sequences, we divide the mixed sequence according to the semantic similarity. As illustrated in Section~\ref{sec:method_global}, the LLMs item embeddings $\mathbf{E}^{LLM}$ store the semantic information of items. Thus, conducting the unsupervised clustering algorithm on $\mathbf{E}^{LLM}$, such as K-means~\cite{na2010research}, can directly lead to the result. The mixed sequence $\tilde{\mathcal{S}}$ will be partitioned into $K$ sub-sequences based on item clusters, denoted as $\tilde{\mathcal{S}}^1, \tilde{\mathcal{S}}^2, \dots \tilde{\mathcal{S}}^K$, where $K$ is the cluster number. 

\subsubsection{\textbf{Profiling}}

In this step, we first adopt LLMs to hierarchically summarize user's preferences. At the \textbf{first} level, we summarize $K$ user preferences based on $K$ sub-sequences. The prompt template is shown as follows.
\begin{tcolorbox}[
        colframe=gray,
        width=1\linewidth,
        arc=1mm, 
        auto outer arc,
        title={Sub-sequence Preference Template},
        breakable,]
        
        Assume you are a consumer who is shopping online. You have shown interest in the following commodities: \\
        \ul{<Item Title 1>; \\
        <Item Title 2>; \\ 
        $\cdots$}. \\
        The commodities are segmented by `\textbackslash n'. \\
        Please conclude it not beyond 50 words. Do not only evaluate one specific commodity but illustrate the interests overall.
        
\end{tcolorbox}

\noindent The words underlined mean the user's interaction history, where the item is represented by its textual title. 
Let $\{P^k\}_{k=1}^K$ denote the textual preferences from $K$ sub-sequences summarized by LLMs. At the \textbf{second} profiling level, we organize $\{P^k\}_{k=1}^K$ into the prompt to get the user's overall preferences for $\tilde{\mathcal{S}}$.

\begin{tcolorbox}[
        colframe=gray,
        width=1\linewidth,
        arc=1mm, 
        auto outer arc,
        title={Overall Preference Template},
        breakable,]
        
        Assume you are a consumer and there are preference demonstrations from several aspects as follows: \\
        \ul{<Summary 1>; \\
        $\cdots$; \\
        <Summary K>.} \\ 
        Please illustrate your preference with fewer than 100 words
        
\end{tcolorbox}

\noindent The sentences underlined in the prompt are the summaries of each sub-sequence, \ie $\{P^k\}_{k=1}^K$. The output of LLMs is the overall cross-domain preference for users, denoted as $\tilde{\mathcal{P}}$.

\subsubsection{\textbf{Alignment}}

To inject the semantic preference contained in texts $\tilde{\mathcal{P}}$ into the LLM4CDSR model, we propose an alignment training method. First, similar to the procedure for item embeddings in Section~\ref{sec:method_global}, the texts $\tilde{\mathcal{P}}$ are converted into embeddings by LLMs, denoted as $\tilde{\mathbf{P}}^{LLM}$. 
As we expect the summarized text can conclude the cross-domain user preference, $\tilde{\mathbf{P}}^{LLM}$ should be similar to the global preference $\tilde{\mathbf{u}}$. Thus, we propose to utilize contrastive learning to align the pair of preference representations, \ie $\tilde{\mathbf{P}}^{LLM}$ and $\tilde{\mathbf{u}}$. 
Specifically, we adopt a two-layer multilayer perception $g(\cdot)$ to shrink the size of $\tilde{\mathbf{P}}^{LLM}$ to $d$, denoted as $\tilde{\mathbf{p}}=g(\tilde{\mathbf{P}}^{LLM})$. Then, an alignment contrastive loss is designed as follows:
\vspace{-1mm}
\begin{equation} \label{eq:align}
    \mathcal{L}^1_{align}=-\frac{1}{Z} \sum_{i=1}^{Z} \log \frac{\exp({\rm cos}(\tilde{\mathbf{u}}_i, \tilde{\mathbf{p}}_i) / \tau)}{\sum_{j=1}^Z \mathbb{I}_{[i \neq j]} \exp({\rm cos}(\tilde{\mathbf{u}}_i, \tilde{\mathbf{p}}_j) / \tau)}
\end{equation}
\noindent where $\tau$ is the temperature coefficient. Exchanging the positions of $\tilde{\mathbf{u}}$ and $\tilde{\mathbf{p}}$, we can obtain the other side of contrastive loss $\mathcal{L}^2_{align}$. The final loss for alignment is $\mathcal{L}_{profile}=\mathcal{L}^1_{align}+\mathcal{L}^2_{align}$.

\let\oldnl\nl
\newcommand{\nonl}{\renewcommand{\nl}{\let\nl\oldnl}}
\begin{algorithm}[!t]
\caption{Training and inference process of LLM4CDSR} \label{alg:train}
\raggedright

\begin{algorithmic} [1]
    \State Indicate the type of LLMs for usage.
    \State Indicate the number of item clusters $K$. 
    \State Indicate weights of regularization and alignment loss, $\alpha$ and $\beta$.
\end{algorithmic}

\textbf{Training Process} 
\setcounter{algorithm}{2}
\begin{algorithmic} [1]
    \makeatletter
    \setcounter{ALG@line}{3}
    \State Derive the LLMs item embedding $\mathbf{E}^{LLM}$.
    \State Derive the LLMs user profiles $\{\tilde{\mathcal{P}}_i\}_{i=1}^{\mid \mathcal{U} \mid}$.
    \While {Convergence}
        \For {a batch $Z$ in $\mathcal{U}$}
            \State Get global and local preferences ($\tilde{\mathbf{u}}$, $\mathbf{u}_A$, $\mathbf{u}_B$) by Eq.~\eqref{eq:user}.
            \State Derive the loss $\mathcal{L}_{SRS}^A$ and $\mathcal{L}_{SRS}^B$ for SRS by Eq.~\eqref{eq:loss_srs}.
            \State Derive the loss $\mathcal{L}_{reg}$ for regularization by Eq.~\eqref{eq:reg}.
            \State Derive the loss $\mathcal{L}_{profile}$ for LLMs profile by Eq.~\eqref{eq:align}.
            \State Calculate total loss by Eq.~\eqref{eq:loss_all} and update the parameters of LLM4CDSR except for $\mathbf{E}^{LLM}$ and $\{\tilde{\mathbf{P}}^{LLM}_i\}_{i=1}^{\mid \mathcal{U} \mid}$.
        \EndFor
    \EndWhile
\end{algorithmic}

\textbf{Inference Process}
\setcounter{algorithm}{13}
\begin{algorithmic} [1]
    \makeatletter
    \setcounter{ALG@line}{14}
    \State Load $\mathbf{E}^{LLM}$ and adapter to get all global item embeddings, denoted as $\tilde{\mathbf{E}}\in\mathbb{R}^{\mid \mathcal{U}\mid \times d}$.
    \If {Serve for Domain $A$}
        \State Input $\tilde{\mathcal{S}}$ and $\mathcal{S}^A$ to calculate the probability by Eq.~\eqref{eq:predict}.
    \Else
        \State Input $\tilde{\mathcal{S}}$ and $\mathcal{S}^B$ to calculate the probability.
    \EndIf

\end{algorithmic}
\end{algorithm}
\vspace{-1mm}

\subsection{Training and Inference} \label{sec:method_train}

We refer to the training and inference processes in this sub-section. In terms of training LLM4CDSR, the loss for SRS, regularization and alignment are weighted summed to optimize the whole framework.
\begin{equation}    \label{eq:loss_all}
    \mathcal{L} = (\mathcal{L}^{A}_{SRS} + \mathcal{L}^{B}_{SRS}) + \alpha \cdot \mathcal{L}_{reg} + \beta \cdot \mathcal{L}_{profile}
\end{equation}
\noindent where $\alpha$ and $\beta$ are hyper-parameters. Note that the two sets of LLMs embeddings $\mathbf{E}^{LLM}$ and $\{\tilde{\mathbf{P}}^{LLM}_i\}_{i=1}^{\mid \mathcal{U} \mid}$ are frozen while training.

For clarity, we conclude all the procedures in Algorithm~\ref{alg:train}. First, we indicate the type of LLMs and hyper-parameters for LLM4CDSR (lines 1-3). Then, to build the semantic connection between items and capture the user's semantic preferences, we prompt LLMs to generate LLMs item embeddings and profiles (lines 4-5). After the preparation, the loss used for optimization is calculated, and the parameters of LLM4CDSR get updated (lines 8-12). As for the inference process, we first derive transformed global embeddings for all items to shrink the embedding layer (line 15). Besides, the hierarchical LLMs profile module is eliminated during serving, due to no need for the alignment loss. Finally, based on the target domain, various interaction sequences are input to LLM4CDSR to get the recommendations (lines 16-20).

%% file: 4Experiment.tex
\section{Experiment}

In this section, we have investigated several \textbf{Research Questions} (\textbf{RQ}) by the experimental results.
\begin{itemize}[leftmargin=*]
    \item \textbf{RQ1}: How does the proposed LLM4CDSR perform compared with existing CDSR and LLM-based SRS models?
    \item \textbf{RQ2}: Is each component designed for LLM4CDSR effective?
    \item \textbf{RQ3}: Can our LLM4CDSR alleviate the overlap dilemma?
    \item \textbf{RQ4}: How do the hyper-parameters affect the LLM4CDSR?
    \item \textbf{RQ5}: Is the LLM4CDSR efficient for inference?
    \item \textbf{RQ6}: Can LLM4CDSR enhance various SRS models?
\end{itemize}

\subsection{Experimental Settings}

\subsubsection{\textbf{Dataset}}    \label{sec:exp-data}
In this paper, we experiment with three public datasets, \ie \textbf{Cloth-Sport}, \textbf{Electronic-Phone} and \textbf{Book-Movie}. The first two datasets originate from sub-categories of the Amazon dataset\footnote{https://cseweb.ucsd.edu/\textasciitilde jmcauley/datasets/amazon\_v2/}, which collect the reviews for commodities on an e-commerce platform. Book-Movie is from the Douban dataset\footnote{https://github.com/fengzhu1/GA-DTCDR/tree/main/Data}, including users' comments on books, movies and music on a Chinese review website. Since we research the implicit feedback for CDSR, all the reviews are considered as positive samples.

For the preprocessing, we first filter out the users who have fewer than $5$ interactions and the items consumed fewer than $3$ times for each domain to avoid the cold-start problem. The records in all domains are organized by timestamp to derive mixed sequences for each user. For training and testing, we split out the penultimate interaction $v_{n_u-1}$ as the validation and the last one $v_{n_u}$ as the test. It is worth noting that, regardless of the domain, only the last item of the mixed sequence $\tilde{\mathcal{S}}$ is used as the test to avoid leakage.
Besides, to further investigate the overlap dilemma, we adjust the overlap ratio in Section~\ref{sec:exp_ratio}. Specifically, we transform an overlap user into a non-overlap one by deleting the interactions of one domain from his or her interaction sequence. By repeating such a process, we can adjust the overlap ratio to $75\%$, $50\%$ and $25\%$.
After preprocessing, the statistical characteristics of the dataset are revealed in Table~\ref{tab:exp-dataset}.

\begin{table}[!t]
\tabcolsep=0.08cm 
\centering
\caption{The statistics of the datasets.}
\resizebox{1\linewidth}{!}{
\begin{tabular}{c|ccc|cc}
\toprule
\textbf{Dataset} & \textbf{\# Users} & \textbf{\# Items} & \textbf{Sparsity} & \textbf{Overlap} & \textbf{Avg.len} \\ 
\midrule
Cloth & 9,933 & 3,278 & 99.70\% & \multirow{2}{*}{3,962} & \multirow{2}{*}{10.71} \\
Sport & 4,263 & 1,021 & 99.04\% &  &  \\ 
\midrule
Electronic & 20,728 & 10,492 & 99.93\% & \multirow{2}{*}{11,698} & \multirow{2}{*}{8.30} \\
Phone & 11,762 & 2,246 & 99.88\% &  &  \\ 
\midrule
Book & 1,381 & 12,426 & 99.88\% & \multirow{2}{*}{1,265} & \multirow{2}{*}{68.30} \\
Movie & 2,213 & 16,537 & 99.92\% &  &  \\ 
\bottomrule
\end{tabular}
}
\vspace{-3mm}
\label{tab:exp-dataset}
\end{table}

\subsubsection{\textbf{Baselines}}
To validate the effectiveness of LLM4CDSR, we compare it with three groups of up-to-date baselines.

\noindent \textbf{i) Single-domain Sequential Recommendation} (SDSR): This category mainly focuses on fabricating neural architectures to capture the user's fine-grained preferences or the loss function to alleviate the data sparsity problem. However, they ignore the supplementary data from other domains or extra semantic information.
\begin{itemize}[leftmargin=*]
    \item \textbf{GRU4Rec}~\cite{hidasi2015session}. It proposes to adopt recurrent neural networks to model the user's dynamic preferences.
    \item \textbf{Bert4Rec}~\cite{sun2019bert4rec}. Following the architecture and training pattern of Bert~\cite{kenton2019bert}, Bert4Rec adopts bi-directional self-attention layers and specifies a close task for the SRS task.
    \item \textbf{SASRec}~\cite{kang2018self}. Different from Bert4Rec, SASRec applies uni-directional self-attention layers to capture the causal relationships contained in the behavior sequence.
    \item \textbf{DuoRec}~\cite{qiu2022contrastive}. To alleviate the data sparsity problem, DuoRec proposes augmentation methods with contrastive learning loss to refine the item and sequence representations.
\end{itemize}

\noindent \textbf{ii) Cross-domain Sequential Recommendation} (CDSR): The CDSR works utilize the interactions from various domains to address the data sparsity issue. 
\begin{itemize}[leftmargin=*]
    \item \textbf{C2DSR}~\cite{cao2022contrastive}. This work utilizes graph neural networks to learn the relationships between items of various domains.
    \item \textbf{TriCDR}~\cite{ma2024triple}. TriCDR captures the user's fine-grained interests from behavior sequences by the designed contrastive objectives.
    \item \textbf{SyNCRec}~\cite{park2024pacer}. SyNCRec mainly targets the negative transfers between domains, designing a mixture-of-expert framework and a gradient stop mechanism to benefit all domains.
    \item \textbf{AMID}~\cite{xu2024rethinking}. To face the overlap dilemma, AMID fabricates a multi-interest information module to learn cross-domain preferences for both overlap and non-overlap users.
\end{itemize}

\noindent \textbf{iii) LLM-based Sequential Recommendation}: Recently, many LLM-based SRS works have emerged, integrating the semantic information derived from LLMs to enhance SRS. However, they are for the single-domain situation, except for URLLM~\cite{shen2024exploring}.
\begin{itemize}[leftmargin=*]
    \item \textbf{SAID}~\cite{hu2024enhancing}. The main idea of SAID lies in injecting semantic information into identity embeddings by fine-tuning LLMs. The derived embeddings initialize the embedding layer of SDSR.
    \item \textbf{TSLRec}~\cite{liu2024practice}. To combine semantic and collaborative signals, TSLRec devises a category prediction task to fine-tune LLMs.
    \item \textbf{LLM-ESR}~\cite{liu2024llm}. Compared to SAID and TSLRec, LLM-ESR highlights LLMs embedding utilization. It proposes a dual-view framework with self-distillation to enhance long-tail users and items.
    \item \textbf{URLLM}~\cite{shen2024exploring}. URLLM is the only work that adopts LLMs to the CDSR task. It ensembles the traditional deep learning models and LLMs for final recommendations.
\end{itemize}

\subsubsection{\textbf{Implementation Details}}
We implement LLM4CDSR and all baselines on an AMD EPYC 7543 32-core platform integrated with RTX 3090 24GB GPUs. The necessary software includes CUDA 12.0, Python 3.9.16, and PyTorch 2.3.1. 
For the two Amazon datasets, the LLMs embeddings are derived by ``text-ada-embedding-002'' API, while the LLMs used to summarize the user's preferences are ``gpt-3.5-turbo-0125'' API\footnote{https://platform.openai.com/docs/models}. By comparison, since Douban is a Chinese dataset, we adopt ``Embedding-3'' and ``GLM-4-Flash'' APIs\footnote{https://bigmodel.cn/dev/api/normal-model/glm-4} for it.
As for the architecture of LLM4CDSR, the number of self-attention layers is $2$, while the dimension is fixed to $128$. During the training, we set the batch size as $128$ and the learning rate as $0.01$ for all datasets. Besides, Adam is used as the optimizer with an early stop strategy. 
The number of item clusters $K$ is fixed to $10$. The temperatures $\gamma$ and $\tau$ are simply set as $1$, because they do not affect the performance evidently. Then, the hyper-parameter $\alpha$ is searched from $\{0.01,0.05,0.1,0.5,1\}$, while $\beta$ from $\{0.1,0.5,1,5,10\}$ for all datsets.

\subsubsection{\textbf{Evaluation Metrics}}
Following previous works~\cite{kang2018self,cao2022contrastive}, we adopt commonly used Top-$k$ metrics, \ie \textit{Hit Rate} (\textbf{H@k}) and \textit{Normalized Discounted Cumulative Gain} (\textbf{N@k}). Specifically, we set $k$ as $10$ in this paper. To guarantee the robustness of the results, we show the average values of $3$ runs with random seeds $\{42,43,44\}$.

\begin{table*}[!t]
\tabcolsep=0.1cm 
\centering
\caption{The overall results of competing baselines and LLM4CDSR on three public datasets. The boldface refers to the highest score and the underline indicates the best result of the methods. ``\textbf{{\Large *}}'' indicates the statistically significant improvements (\ie two-sided t-test with $p<0.05$) over the best baseline.}
\resizebox{1\textwidth}{!}{
\begin{tabular}{cc|cccc|cccc|cccc}
\toprule
\multicolumn{2}{c|}{\multirow{3}{*}{\textbf{Model}}} & \multicolumn{4}{c|}{\textbf{Amazon}} & \multicolumn{4}{c|}{\textbf{Amazon}} & \multicolumn{4}{c}{\textbf{Douban}} \\ 
\cmidrule{3-14} 
\multicolumn{2}{c|}{} & \multicolumn{2}{c}{\textbf{Cloth}} & \multicolumn{2}{c|}{\textbf{Sport}} & \multicolumn{2}{c}{\textbf{Electronic}} & \multicolumn{2}{c|}{\textbf{Phone}} & \multicolumn{2}{c}{\textbf{Book}} & \multicolumn{2}{c}{\textbf{Movie}} \\ 
\cmidrule{3-14} 
\multicolumn{2}{c|}{} & \textbf{H@10} & \textbf{N@10} & \textbf{H@10} & \textbf{N@10} & \textbf{H@10} & \textbf{N@10} & \textbf{H@10} & \textbf{N@10} & \textbf{H@10} & \textbf{N@10} & \textbf{H@10} & \textbf{N@10} \\ 
\midrule
\multirow{4}{*}{\textbf{SDSR}} 
& GRU4Rec & 0.5806 & 0.5252 & 0.5736 & 0.4779 & 0.3225 & 0.1869 & 0.3101 & 0.1772 & 0.2101 & 0.1286 & 0.7211 & 0.6386 \\
& Bert4Rec & 0.6531 & 0.5720 & 0.5350 & 0.4514 & 0.3332 & 0.1903 & 0.3001 & 0.1714 & 0.2693 & 0.1755 & 0.7795 & 0.6144 \\
& SASRec & 0.7057 & 0.6543 & 0.5964 & 0.4900 & 0.3586 & 0.2234 & 0.3292 & 0.2011 & 0.2684 & 0.1740 & 0.7839 & 0.6181 \\
& DuoRec & 0.7116 & 0.6528 & 0.6254 & 0.5782 & 0.3891 & 0.2442 & 0.3287 & 0.2000 & 0.2856 & 0.1885 & 0.8078 & 0.6507 \\ 
\midrule
\multirow{4}{*}{\textbf{CDSR}} 
& C2DSR & 0.7251 & 0.6760 & 0.6072 & 0.5703 & 0.4548 & 0.3196 & 0.2846 & 0.1735 & 0.3226 & 0.2368 & 0.7925 & 0.6285 \\
& TriCDR & 0.7301 & 0.6789 & 0.6157 & 0.5779 & 0.4543 & \underline{0.3232} & 0.2624 & 0.1601 & 0.3333 & 0.2405 & 0.7909 & 0.6263 \\
& SyNCRec & 0.7359 & 0.6642 & 0.6262 & 0.5664 & 0.4391 & 0.3093 & 0.3359 & 0.2047 & 0.3818 & 0.2601 & 0.7879 & 0.6176 \\
& AMID & \underline{0.7475} & \underline{0.6814} & 0.6377 & 0.5867 & \underline{0.4683} & 0.3211 & 0.3362 & 0.2116 & \underline{0.4488} & \underline{0.2851} & 0.8199 & 0.6453 \\ 
\midrule
\multirow{3}{*}{\textbf{LLM-based}} 
& SAID & 0.7394 & 0.6755 & 0.6242 & 0.5727 & 0.4508 & 0.2769 & 0.3319 & 0.2032 & 0.3515 & 0.2425 & 0.8354 & 0.6638 \\
& TSLRec & 0.7183 & 0.6605 & 0.6192 & 0.5778 & 0.4048 & 0.2467 & 0.3411 & 0.2107 & 0.2951 & 0.1823 & 0.7867 & 0.6240 \\
& LLM-ESR & 0.7472 & 0.6810 & \underline{0.6436} & \underline{0.5965} & 0.4599 & 0.2828 & \underline{0.3412} & \underline{0.2127} & 0.4183 & 0.2705 & \underline{0.8319} & 0.6676 \\
& URLLM & 0.7327 & 0.6785 & 0.6384 & 0.5813 & 0.4448 & 0.3133 & 0.3324 & 0.2075 & 0.3582 & 0.2510 & 0.8230 & \underline{0.6919} \\ 
\midrule
\multirow{2}{*}{\textbf{Ours}} 
& \cellcolor{gray!20}\textbf{LLM4CDSR} & \cellcolor{gray!20}\textbf{0.8018}* & \cellcolor{gray!20}\textbf{0.7316}* & \cellcolor{gray!20}\textbf{0.7046}* & \cellcolor{gray!20}\textbf{0.6312}* & \cellcolor{gray!20}\textbf{0.4886}* & \cellcolor{gray!20}\textbf{0.3342}* & \cellcolor{gray!20}\textbf{0.3540}* & \cellcolor{gray!20}\textbf{0.2249}* & \cellcolor{gray!20}\textbf{0.5912}* & \cellcolor{gray!20}\textbf{0.3857}* & \cellcolor{gray!20}\textbf{0.8791}* & \cellcolor{gray!20}\textbf{0.7820}* \\
& Impr (\%) & 7.26 & 7.37 & 9.48 & 5.82 & 4.33 & 3.40 & 3.75 & 5.74 & 31.73 & 35.29 & 5.67 & 13.02 \\ 
\bottomrule
\end{tabular}
}
\label{tab:exp-overall}
\vspace{-3mm}
\end{table*}

\subsection{Overall Performance (RQ1)}

We show the overall performance of our LLM4CDSR and competing baselines in Table~\ref{tab:exp-overall} to answer \textbf{RQ1}. At a glance, the proposed LLM4CDSR outperforms all SDSR, CDSR and LLM-based baselines on three datasets consistently, revealing the effectiveness of our method. Then, a more detailed analysis of the results will be given.

Observing the performance of SDSR baselines, they often lag behind other groups broadly. The reasons lie in the sparsity issue and lack of semantic information. Nevertheless, we find that DuoRec performs close to several CDSR and LLM-based models, because it utilizes contrastive learning to alleviate the sparsity problem. Compared to the SDSR group, CDSR models perform much better, especially in the domains with fewer interactions, \eg Sport and Book domains. This phenomenon illustrates that CDSR can address the data sparsity problem by absorbing more data from various domains. Furthermore, C2DSR and TriCDR often lag behind the other CDSR competitors, which validates that they suffer from the overlap dilemma and transition complexity severely.
AMID significantly leads the performance across CDSR baselines, because it designs a special information transfer module to address the overlap dilemma. However, our LLM4CDSR is still superior to it evidently, indicating that LLMs can better bridge the connections between items of different domains and capture users' overall preferences. 

LLM-based SRS models have shown comparable performance due to the extra semantic information from LLMs. Specifically, SAID and LLM-ESR perform better, since they actually prompt LLMs to understand items by attributes. In contrast, TSLRec mainly utilizes the reasoning abilities of LLMs, only adopting item identities. Nevertheless, they are still trapped in the data of a single domain, leading to compromised performance compared with LLM4CDSR.
Though URLLM combines CDSR and LLMs, it prompts LLMs to generate recommendations directly, limited by out-of-corpus issues~\cite{bao2023bi} and lack of collaborative signals. In contrast, LLM4CDSR still adopts the discriminative RS model as the backbone, remitting these two problems caused by LLMs themselves.

\begin{table}[!t]
\centering
\caption{The experiments for the ablation study.}
\begin{tabular}{c|cccc}
\toprule
\multirow{2}{*}{\textbf{Domain}} & \multicolumn{2}{c}{\textbf{Cloth}} & \multicolumn{2}{c}{\textbf{Sport}} \\ 
\cmidrule{2-5} 
 & \multicolumn{1}{c}{\textbf{H@10}} & \multicolumn{1}{c}{\textbf{N@10}} & \multicolumn{1}{c}{\textbf{H@10}} & \multicolumn{1}{c}{\textbf{N@10}} \\ 
\midrule
\cellcolor{gray!20}\textbf{LLM4CDSR} & \cellcolor{gray!20}\textbf{0.8018} & \cellcolor{gray!20}\textbf{0.7316} & \cellcolor{gray!20}\textbf{0.7046} & \cellcolor{gray!20}\textbf{0.6312} \\
\textit{w/o} Unified & 0.7402 & 0.6361 & 0.5992 & 0.5368 \\
\textit{w/o} Profile & 0.7888 & 0.7200 & 0.6842 & 0.6256 \\
\textit{w/o} Reg & 0.7807 & 0.7154 & 0.6830 & 0.6162 \\
\textit{w/o} Cluster & 0.7908 & 0.7206 & 0.6896 & 0.6195 \\
\textit{w/o} Init & 0.7439 & 0.6822 & 0.6829 & 0.6208 \\ 
\midrule
\multirow{2}{*}{\textbf{Domain}} & \multicolumn{2}{c}{\textbf{Book}} & \multicolumn{2}{c}{\textbf{Movie}} \\ 
\cmidrule{2-5} 
 & \multicolumn{1}{c}{\textbf{H@10}} & \multicolumn{1}{c}{\textbf{N@10}} & \multicolumn{1}{c}{\textbf{H@10}} & \multicolumn{1}{c}{\textbf{N@10}} \\ 
\midrule
\cellcolor{gray!20}\textbf{LLM4CDSR} & \cellcolor{gray!20}\textbf{0.5912} & \cellcolor{gray!20}\textbf{0.3857} & \cellcolor{gray!20}0.8791 & \cellcolor{gray!20}\textbf{0.7820} \\
\textit{w/o} Unified & 0.4728 & 0.3068 & 0.8385 & 0.6836 \\
\textit{w/o} Profile & 0.5387 & 0.3669 & 0.8712 & 0.7780 \\
\textit{w/o} Reg & 0.5661 & 0.3843 & 0.8698 & 0.7746 \\
\textit{w/o} Cluster & 0.5761 & 0.3681 & \textbf{0.8809} & 0.7787 \\
\textit{w/o} Init & 0.5352 & 0.3773 & 0.8710 & 0.7510 \\ 
\bottomrule
\end{tabular}
\label{tab:exp-ablation}
\vspace{-3mm}
\end{table}

\subsection{Ablation Study (RQ2)}

To investigate whether each designed component contributes to LLM4CDSR, we experiment with several variants. The results are shown in Table~\ref{tab:exp-ablation}. First, we remove the proposed unified LLMs representation module by training both local and global embedding layers from scratch, denoted as \textbf{\textit{w/o} Unified}. This variant shows an obvious performance drop in both domains, highlighting the effects of semantic connections for the overlap dilemma. \textbf{\textit{w/o} Profile} eliminates the alignment loss $\mathcal{L}_{profile}$, which means removing the hierarchical LLMs profile module directly. We find this variant is inferior to LLM4CDSR, especially for Cloth and Book, illustrating transition complexity affects weak domains more severely.

Next, to explore the effectiveness of the contrastive regularization, we remove the loss $\mathcal{L}_{reg}$ as \textbf{\textit{w/o} Reg}. The results reveal that dropping the designed regularization will bring performance decreases consistently, which validates that contrastive regularization helps learn domain-specific information. \textbf{\textit{w/o} Cluster} represents the variant that generates semantic summary by LLMs without partitioning the behavior histories into several sub-sequences. The overall performance drops prove the negative impacts caused by lengthy prompts. Finally, we do not initialize the local embedding layers by the LLMs item embeddings, \ie \textbf{\textit{w/o} Init}. This variant underperforms the original LLM4CDSR evidently, indicating the designed initialization can help alleviate difficulty in training. As the response to \textbf{RQ2}, all designs for LLM4CDSR make effects.

\subsection{Overlap Study (RQ3)}  \label{sec:exp_ratio}

\begin{figure}[!t]
\centering
\includegraphics[width=1\linewidth]{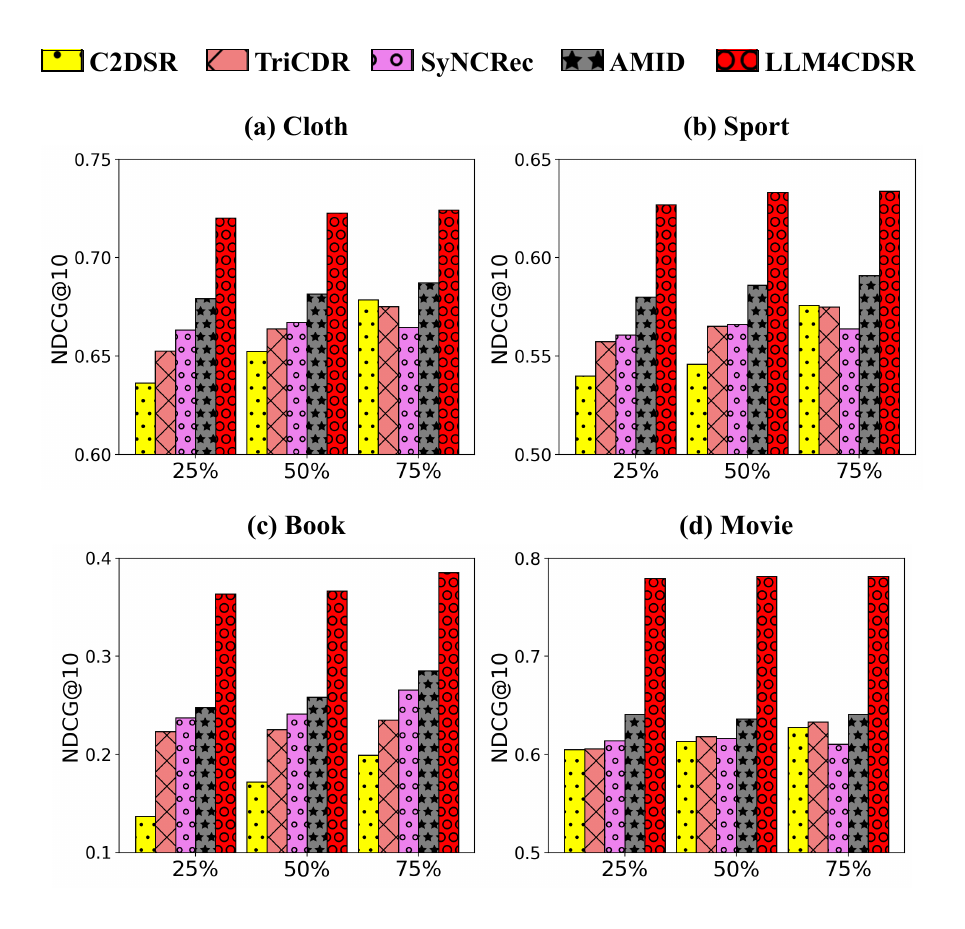}
\caption{The performance comparison under various ratios.}
\label{fig:exp-ratio}
\vspace{-3mm}
\end{figure}

As introduced in Section~\ref{sec:intro}, the existing CDSR models suffer from the overlap dilemma. To verify whether LLM4CDSR can alleviate this problem, \ie \textbf{RQ3}, we further test our model and CDSR baselines under various overlap ratios. The process of altering the ratio is illustrated in Section~\ref{sec:exp-data}. As shown in Figure~\ref{fig:exp-ratio}, the proposed LLM4CDSR outperforms all competitors under any overlap ratio, which indicates it can address the overlap dilemma effectively. Moreover, we find that the performance of TriCDR and C2DSR drops sharply when the ratio decreases, because they totally rely on the users who interact with both domains. Besides, though SyNCRec and AMID have addressed the overlap dilemma from the perspectives of negative transfer and user modeling, they are still affected by the ratio largely and inferior to LLM4CDSR. This comparison highlights the merit of the semantic connections built by LLMs.

\begin{figure}[!t]
\centering
\includegraphics[width=1\linewidth]{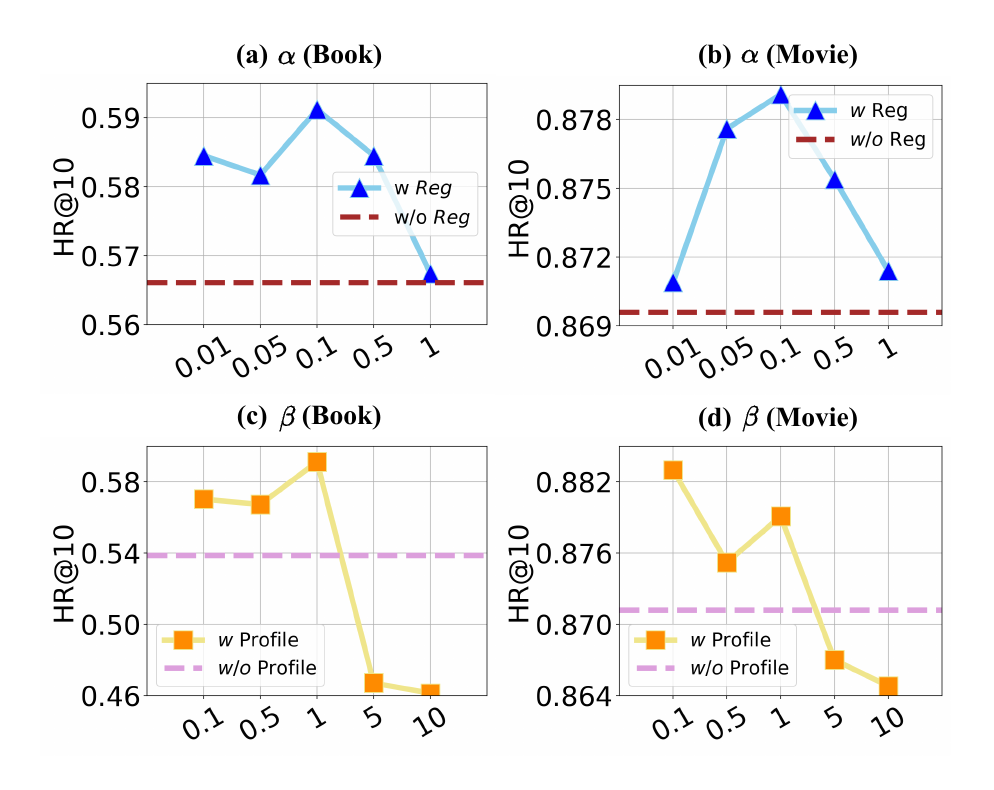}
\caption{The results of hyper-parameter experiments on the Douban dataset.}
\label{fig:exp-hyper}
\vspace{-4mm}
\end{figure}



\subsection{Hyper-parameter Analysis (RQ4)}

In terms of \textbf{RQ4}, we vary the weights for regularization and alignment loss, \ie $\alpha$ and $\beta$, and show the performance trends in Figure~\ref{fig:exp-hyper}. Observing the results of $\alpha$, we find that the hit rate presents rising first and dropping then within the range of $[0.01, 1]$. Such a phenomenon illustrates that the designed regularization is effective in capturing domain-specific relationships, while overemphasis on contrastive loss leads to optimization difficulty. As for the profile alignment loss, the results indicate the best value for $\beta$ is $1$ broadly. When $\beta$ increases from $1$ to $10$, the performance decreases because of the possible noise contained in the LLMs-generated profile. By comparison, the trends for Book and Movie domains differ with $\beta$ rising from $0.1$ to $0.01$. The reason lies in that the Movie domain owns more data, where the effects of noise oppress the benefits of semantic preferences. For the sparser Book domain, a larger $\beta$ is more beneficial due to supplementing overall preferences by LLMs.

\subsection{Efficiency Study (RQ5)}

\begin{table*}[!t]
\centering
\caption{The efficiency comparison on the Douban dataset. The model size and time for inferring all users are shown.}
\label{tab:exp-efficiency}
\begin{tabular}{c|c|cccc|cc|c}
\toprule
\textbf{Models}  & \textbf{SASRec} & \textbf{C2DSR} & \textbf{TriCDR} & \textbf{SyNCRec} & \textbf{AMID} & \textbf{LLM-ESR} & \textbf{URLLM} & \textbf{LLM4CDSR} \\ 
\midrule
\textbf{Parameter (\# M)}  & \textbf{3.99} & 7.81 & \underline{7.72} & 9.73 & \textbf{3.99} & 23.48 & 6335.25 & \cellcolor{gray!20}\underline{7.72}         \\
\textbf{Time (\# s)} & 8.34 & \underline{6.19} & \textbf{5.89} & 7.41 & 6.82 & 11.15 & 7065.02 & \cellcolor{gray!20}\textbf{5.89}     \\ 
\bottomrule
\end{tabular}
\vspace{-3mm}
\end{table*}

To respond to the \textbf{RQ5}, we conduct efficiency experiments, as shown in Table~\ref{tab:exp-efficiency}. The ``Parameter'' refers to the size of the model, while ``Time'' is the inference time. As for the model size, SASRec and AMID are the lightest models, because they do not model the mixed sequence explicitly. As for LLM4CDSR, it has the same architecture as TriCDR during inference.
Besides, LLM4CDSR and TriCDR outperform other models on latency. CDSR models can serve one user for both domains simultaneously, but SASRec has more inference time due to twice the inference. 
In conclusion, our LLM4CDSR achieves the best in recommending accuracy and latency while being comparable in model size.

\subsection{Generality Study (RQ6)}

We compare the performance of LLMCDSR and baselines integrated with different SRS models in Table~\ref{tab:exp-compat} to answer \textbf{RQ6}. The results show that LLM4CDSR outperforms the baselines under both GRU4Rec and Bert4Rec models. Besides, the overall results of AMID, LLM-ESR and LLM4CDSR in Table~\ref{tab:exp-overall} can be regarded as their performance with SASRec as the backbone. All these comparisons verify the better generality of our LLM4CDSR, indicating its further potential in real-world applications.

\begin{table}[!t]
\centering
\caption{The experiments with various SRS on the Douban Dataset. ``-'' indicates the original SRS models.}
\label{tab:exp-compat}
\tabcolsep=0.05cm 
\resizebox{1\linewidth}{!}{
\begin{tabular}{c|cccc|cccc}
\toprule
\multirow{3}{*}{\textbf{Model}} & \multicolumn{4}{c|}{\textbf{GRU4Rec}} & \multicolumn{4}{c}{\textbf{Bert4Rec}}  \\
\cmidrule{2-9} 
& \multicolumn{2}{c}{\textbf{Book}} & \multicolumn{2}{c|}{\textbf{Movie}} & \multicolumn{2}{c}{\textbf{Book}} & \multicolumn{2}{c}{\textbf{Movie}} \\ 
\cmidrule{2-9} 
& \textbf{H@10} & \textbf{N@10} & \textbf{H@10} & \textbf{N@10} & \textbf{H@10} & \textbf{N@10} & \textbf{H@10} & \textbf{N@10} \\ 
\midrule
-  & 0.2101 & 0.1286 & 0.7211 & 0.6386 & 0.2693 & 0.1755 & 0.7795 & 0.6144           \\
AMID & 0.2521 & 0.1574 & 0.7640 & 0.6400 & 0.4550 & 0.2646 & 0.8052 & 0.6351           \\
LLM-ESR & 0.2668 & 0.1579 & 0.7773 & 0.6423 & 0.4298 & 0.2922 & 0.8313 & 0.6780           \\
\cellcolor{gray!20}\textbf{LLM4CDSR} & \cellcolor{gray!20}\textbf{0.2865} & \cellcolor{gray!20}\textbf{0.1649} & \cellcolor{gray!20}\textbf{0.8037} & \cellcolor{gray!20}\textbf{0.7083} & \cellcolor{gray!20}\textbf{0.5559} & \cellcolor{gray!20}\textbf{0.4123} & \cellcolor{gray!20}\textbf{0.8815} & \cellcolor{gray!20}\textbf{0.7847}           \\ 
\bottomrule
\end{tabular}
}
\vspace{-3mm}
\end{table}

%% file: 5RelatedWork.tex
\section{Related Works}


\noindent \textbf{Cross-domain Sequential Recommendation}.
Recommender systems~\cite{zhao2018deep,zhao2018recommendations,liu2023multi,wang2023multi,wang2023single,liu2024multimodal} become important in life since they can alleviate the problem of information overload.
Recently, CDSR~\cite{chen2024survey} has emerged, absorbing both merits from cross-domain recommendation~\cite{li2022gromov,wang2023plate,li2023hamur,gao2023autotransfer,zhang2024m3oe,jia2024d3,liu2024multifs} and sequential recommendation~\cite{liu2025sigma,liu2023diffuasr,li2023strec,liu2023dirac}.
The key to CDSR lies in filling the distribution gap between various domains. 
From the item aspect, most current studies adopt Graph Neural Networks (GNN)~\cite{wu2022graph} to establish the item connections across domains. As one of the pioneers, C2DSR~\cite{cao2022contrastive} proposes to build an interaction graph based on the co-occurrences of the items across two domains. 
Following C2DSR, EA-GCL~\cite{wang2023unbiased} and MGCL~\cite{xu2023multi} further design contrastive objectives for graph learning to alleviate density bias and transition difficulty. 
From the user perspective, CDSR works often design more sophisticated architectures to capture global preferences from the mixed behavior sequences. $\pi$-Net~\cite{ma2019pi}, an early work in this line, proposes a cross-domain transfer module to derive hybrid preferences. Recently, TriCDR~\cite{ma2024triple} has focused on modeling mixed sequences, designing triple contrastive tasks to dig into fine-grained global interests. Though existing CDSR works have explored various methods to bridge the domains, they are trapped in a collaborative view. 
In contrast, we propose a semantic view to enhance CDSR by LLMs.

\vspace{1mm}
\noindent \textbf{LLMs for Sequential Recommendation}.
LLMs have been applied to several fields~\cite{xu2024multi,liu2024moelora}, including recommender systems~\cite{wu2024survey,lin2023can,bao2024large,wang2024towards,liu2025llmemb,sun2025llmser,liu2024leader,wang2024llm4msr,fu2023unified}. Relying on powerful abilities in reasoning and understanding~\cite{zhao2023survey}, LLMs can benefit SRS by analyzing users' behaviors and items' attributes. Existing LLMs for SRS works can be categorized into two main lines, \ie LLMs as SRS and LLMs enhancing SRS. 
The first category refers to utilizing LLMs to generate recommendations directly. For example, GOT4Rec~\cite{long2024got4rec} designs chain-of-thoughts to prompt LLMs giving out recommendations. To fill the gap between natural language and recommendation tasks, some propose to fine-tune open-sourced LLMs, \eg TALLRec~\cite{bao2023tallrec} and S-DPO~\cite{chen2024softmax}. 
The other line, \ie LLMs enhancing SRS, often adopts LLMs as item or user encoders~\cite{liu2024llmers}. Since the LLMs embeddings can be cached in advance, they are more practical due to no need for LLMs while serving. SAID~\cite{hu2024enhancing} and TSLRec~\cite{liu2024practice} are two representatives, which use the LLMs embeddings to initialize the embedding layer of SRS. To better combine collaborative and semantic information, LLM-ESR~\cite{liu2024llm} designs a dual-view modeling framework. 
It is worth noting that we are the first to explore how to utilize LLMs to enhance CDSR.

%% file: 6Conclusion.tex
\section{Conclusion}

In this paper, we design an LLMs-enhanced cross-domain sequential recommendation model (LLM4CDSR). To lighten the requests for overlapping users, we utilize LLMs to generate global item embeddings, which build the semantic connections between items across various domains. Then, a hierarchical profiling module is proposed to conclude users' cross-domain preferences from a semantic perspective, alleviating the transition complexity problem. By the comprehensive experiments on three public cross-domain datasets, the LLM4CDSR is verified as effective and efficient.
In the future, we will explore how to fine-tune LLMs to generate embeddings and profiles that are more suitable to the CDSR task.

%% file: 8Acknowledgement.tex
\begin{acks}
    This research was partially supported by 
    Huawei (Huawei Innovation Research Program, Huawei Fellowship), Research Impact Fund (No.R1015-23), and Collaborative Research Fund (No.C1043-24GF),  
    National Natural Science Foundation of China (No.62192781, No.62177038, No.62293551, No.62277042, No.62137002, No.61721002, No.61937001, No.62377038), Project of China Knowledge Centre for Engineering Science and Technology. 
    
\end{acks}